\documentclass[twocolumn,aps]{revtex4}
\usepackage{graphicx}

\begin{document}
\title{Optimization of a Langmuir-Taylor detector for lithium}
\author{R. Delhuille}
\author{A. Miffre}
\author{E. Lavallette}
\author{M. B\"uchner}
\author{C. Rizzo}
\author{G. Tr\'enec}
\author{J. Vigu\'e}
\email{jacques.vigue@irsamc.ups-tlse.fr}
\affiliation{Laboratoire Collisions Agr\'egats R\'eactivit\'e - IRSAMC
\\ Universit\'e Paul Sabatier and CNRS UMR 5589
\\ 118, Route de Narbonne 31062 Toulouse Cedex, France}
\author{H. J. Loesch}
\affiliation{Fakult\"at f\"ur Physik, Universit\"at Bielefeld
\\ Universit\"atsstrasse 25, D-33615 Bielefeld, Germany}
\author{J. P. Gauyacq}
\affiliation{Laboratoire des Collisions Atomiques et Mol\'eculaires
\\ Universit\'e Paris-Sud and CNRS UMR 8625,
\\ B\^atiment 351, Universit\'e Paris-Sud, 91405 Orsay Cedex, France}




\begin{abstract}

This paper describes the construction and
optimization of a Langmuir-Taylor detector for lithium, using a
rhenium ribbon. The absolute detection probability of this very 
sensitive detector is measured and the dependence of this probability with
oxygen pressure and surface temperature is studied. Sources of background signal and their minimization are also discussed in details. And a comparison between our data concerning the response time of the detector and literature values is given. A theoretical analysis has been made: this analysis supports the validity of the Saha-Langmuir law to relate the ionization probability to the work function. Finally, the rapid variations of the work function with oxygen pressure and temperature are explained by a chemical equilibrium model.
\vspace{2 cm}
\end{abstract}

\maketitle

\section{Introduction}

Very sensitive detectors for neutral atoms, which were first
necessary for Rabi experiments (see reviews in ref.
\cite{king56,ramsey56}), still govern the feasibility of many
experiments. Two different detectors fulfill the requirements of
high efficiency and low background:
\begin{itemize}
\item the Langmuir-Taylor detector \cite{langmuir25,taylor29,taylor30}
is based on the surface ionization process. For ground state
atoms, the ionization probability is large when the ionization
potential is low. Therefore, this detector is mostly used with
alkali atoms. For thermal atoms, the detection probability is
almost independent of their velocities. The detection of a few
atoms per second is feasible, if the produced ions are detected
with an electron multiplier.

\item the laser induced fluorescence detector, which is commonly used
in cold atom experiments, has a detection probability close to
$100 \%$ for slow atoms. However, its detection probability
decreases rapidly with the atom velocity and a large efficiency is
very difficult to achieve for thermal atoms. Finally, laser stray
light is usually the dominant source of background and limits the
detection of a very low atomic flux.
\end{itemize}

Therefore, the Langmuir-Taylor detector is probably the best
detector for thermal alkali atomic beams. Optimization of this
detector is more difficult in the case of lithium, because lithium
has the highest ionization potential of the alkali atoms. In the
present paper, we give a detailed description of the detector in
this case. More precisely, after the introduction, we describe the
detector principle and its design in part 2. In part 3, we present
our measurements of the detection probability, which depends on
the degree of oxidation of rhenium surface
\cite{persky68,kawano86}. The various sources of background signal
and their minimization are described in part 4, while, in part 5,
we discuss the detector response time. In part 6, these
observations are rationalized by a theoretical modeling of the
surface ionization process. Two appendices present complementary
information: appendix A discusses the relation between heating
current and rhenium ribbon temperature and appendix B briefly
analyzes literature data concerning the dependence of rhenium work
function with surface oxidation \cite{kawano86,kawano98,kawano00,kawano99}.

All the information needed to optimize this detector is thus
collected and analyzed in the present paper. A recent paper by F.
Stienkemeier et al. \cite{stienkemeier00} has described the
Langmuir-Taylor detector, using a rhenium surface, applied to
detect various atoms (including lithium) attached to helium
droplets. Our measurements and analysis are largely complementary
to those of this work.

\section{Detector principle and design}

\subsection{Detector principle}

Surface ionization of an atom $A$ occurs if the ionization
potential $I$ of the atom is comparable to the work function
$\Phi$ of the metal. The atom is then emitted as a positive ion
$A^+$ with a probability $P_+$ and as a neutral atom with the
probability $(1-P_+)$. The probability $P_+$ is usually assumed to
be given by the Saha-Langmuir law:
\begin{equation}
\label{principle1}
 P_+ = \frac{1}{1 + \frac{g_0}{g_+} \exp(\frac{I-\Phi}{k_B T})}
\end{equation}
\noindent where $g_0$ and $g_+$ are the statistical weights of the
ion and atom ground states (in the case of alkali atoms, $g_0/g_+
=2$). The validity of this law is discussed in part 6. The wire
temperature must be high, typically $1500$ K or larger, not
because of its influence on the ionized fraction $P_+$, but to
reduce the ion residence time on the surface. This residence time
$\tau$ is given by:
\begin{equation}
\label{principle2}
 \tau = \tau_0 \exp\left( E_{ads}/k_BT\right)
\end{equation}
\noindent Here $E_{ads}$ is the ion adsorption energy (typically a
few eV) and $\tau_0$ should be close to the vibrational period of
the ion near the surface (near $10^{-13}$ s, see table 2).
Finally, the ionization probability is independent of the initial
kinetic energy as long as this energy is smaller than or
comparable to the adsorption energy, because the residence time on
the surface is sufficient to insure thermal equilibrium with the
surface.

Because of the high ionization potential of lithium, $I_{Li} =
5.392$ eV, we need a metal with a large work function $\Phi$.
Among refractory metals, the highest values of the work function
are for platinum ($\Phi = 5.77 $ eV), iridium ($\Phi = 5.70 $ eV)
\cite{kaack95}, rhenium ($\Phi = 4.96$ eV) and tungsten ($\Phi=
4.54$ eV) \cite{kawano99}. Moreover, oxidation increases the
tungsten work function ($\Phi\approx 6$ eV following N. F. Ramsey
\cite{ramsey56}, $\Phi\approx5.9-6.2$ eV following H. Pauly and J.
P. Toennies \cite{pauly68}) and a similar effect occurs for
rhenium \cite{persky68,kawano86,zandberg95,kawano99}. In appendix
B, we discuss the dependence of rhenium work function with
temperature and oxygen partial pressure: this information
supports strongly the idea that rhenium is probably the best
material for a lithium detector. The work function is also a
function of the alkali surface coverage
\cite{kaack95,kawano86,zandberg95}. This effect is very important
if one uses a high flux especially when the residence time is long
(i.e. at low temperature) but it is usually negligible for a high
sensitivity detector dedicated to very low atomic flux.

\subsection{Detector design}

\begin{figure}[htb]
\includegraphics{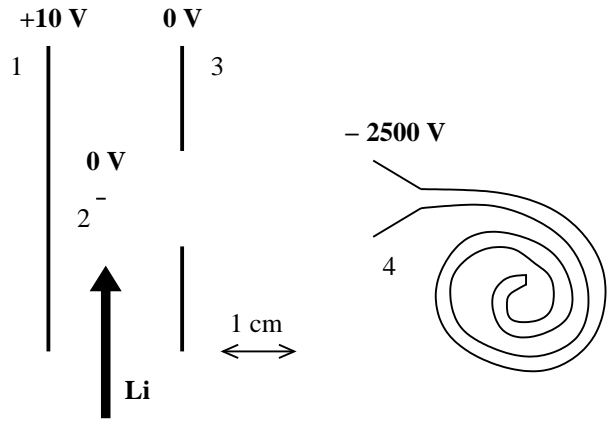}
\caption{\label{design} Schematic drawing of the detector design:
1 repeller plate; 2 endview of rhenium ribbon; 3 electrostatic
lens; 4 channeltron. The arrow gives the scale and a typical
choice of potentials is also indicated.}
\end{figure}

The central part of the detector is a rhenium ribbon. The ribbon
thickness $a$ must be quite small, to facilitate resistive heating
and cleaning (see part 4), while the width $b$ is fixed by
experimental needs. For example, we use commercially available
ribbons with $a= 0.0008 " =20.3$ $\mu m$ and $b=0.040 " = 1016 $
$\mu m$ provided by A. D. Mackay or $a= 30$ $\mu m$ and $b=760$
$\mu m$ provided by Goodfellow. The ribbon is a few centimeters
long. It is heated by circulating a DC current of a few amperes,
necessary to reach the operating temperature $T \simeq 1500$ K. In
appendix A, we discuss the relation between the current and the
ribbon temperature. Higher temperatures ($ T\geq 2200$ K) are
needed during the cleaning process. At $T= 2000$ K, the power per
unit length is about $5$ W/cm and this heat load may induce some
outgassing of the vacuum tank.

The ribbon expansion \cite{goodfellow,rembar} when going from
ordinary temperature to a temperature near $2000$ K is close to
$1\%$. Usually, the ribbon is kept straight by a soft spring, with
an applied force less than $1$ Newton (for data concerning the
mechanical strength of rhenium, see ref.
\cite{goodfellow,rembar}). In our arrangement, the rhenium ribbon
is spot-welded at both ends to tantalum sheets.

The ions emitted by the wire are collected and focussed on the
entrance funnel of a channeltron, by a simple ion optics, as shown
in figure \ref{design}. The ion trajectories have been calculated
with the SIMION software \cite{SIMION} and this calculation serves
to define the various electrical potentials, which are further
optimized by maximizing the ion signal. The repeller plate
potential must not exceed a few volts, otherwise the electrons
emitted by the hot wire may ionize the background gas, thus
contributing to the background signal. A fast counting electronics
is used to convert the channeltron pulses into a signal expressed
in counts per second.

\section{Detection probability}

The principle of our measurement is to use an effusive atomic beam
of lithium. As the theory of such a beam is perfectly understood
\cite{ramsey56}, we can calculate the atomic flux $dN/dt$ reaching
the ribbon:
\begin{equation}
\label{atomflux}
 \frac{dN}{dt} = {\mathcal I} \Delta\Omega
\end{equation}
\noindent where the beam intensity $\mathcal{I}$ is given by
${\mathcal I} = n_{Li} v_{mean} a^*/(4\pi)$ ($n_{Li}$ is the
lithium density in the oven, $v_{mean} = \sqrt{8k_BT/\pi m}$ is
the mean velocity inside the oven, $a^*$ is the area of the oven
exit hole) and $\Delta\Omega$ is the solid angle of the rhenium
wire seen from the oven. In some experiments, we used a
piezoelectric slit, whose width is tunable under vacuum, to reduce
this solid angle, thus verifying the detector linearity with the
atomic flux. Under these conditions, the signal remains in the
linearity domain of the channeltron as long as the lithium oven
temperature does not exceed $673$ K.

Several equations relating the lithium vapor pressure to the
temperature appear in the literature
\cite{ditchburn41,nesmeyanov63}. Nesmeyanov \cite{nesmeyanov63}
considers that the most reliable data covering the $735-915$ K
range are represented by the equation:
\begin{equation}
\label{pressure2}
 log_{10}\left(p_{Li}\right) = 8.012 - 8172/T
\end{equation}
\noindent with the pressure in Torr. We use the perfect gas
approximation to convert pressure to density by $n_{Li} =
p_{Li}/k_BT$.

\begin{figure}[t]
\includegraphics{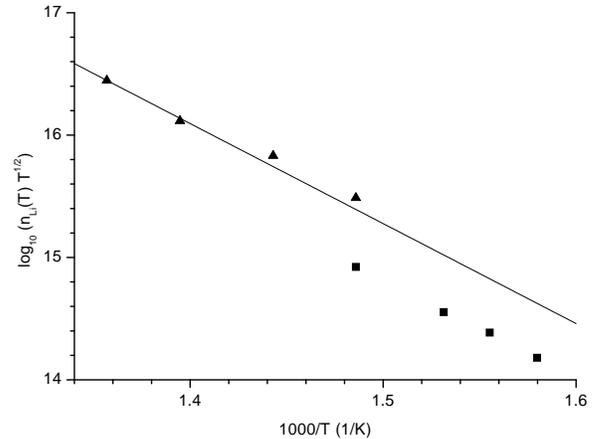}
\caption{\label{efficiency} Beam intensity $\mathcal{I}$ as a
function of the  oven temperature $T$. The plot presents the
logarithm of $\mathcal{I}\sqrt{T}$ versus $1/T$ ($T$ in Kelvin) so
that the theoretical curve deduced from equation (\ref{pressure2})
is a straight line. The triangles represent our absorption
measurements, calibrated at the highest temperature using equation
(\ref{pressure2}). The squares represent the same quantity deduced
from the Langmuir-Taylor detector signal, assuming a $100\%$
efficiency. The absorption data prove that extrapolation of the
pressure law is reasonably accurate. The Langmuir-Taylor detector
signals also support the validity of the extrapolation of this
pressure law. Finally, we deduce from this plot the detection
probability $32\pm 5$ \%.}
\end{figure}

As we must extrapolate this equation, we made a laser absorption
experiment with the same atomic beam to test this extrapolation.
Because of laser saturation and optical pumping, the theoretical
description of laser absorption is not very simple and we use it
here just as a relative measurement. The atomic density $n$ in the
beam is proportional to the absorption:

$$n \propto \ln(I_t/I_0) \approx (I_0 - I_t)/I_0$$

\noindent where $I_0$ and $I_t$ are the incident and transmitted
laser intensities. This experiment was made in the temperature
range $673-777$ K. To calibrate this measurement, we use the vapor
pressure at the highest temperature of our study. The results of
the absorption experiment are shown in figure \ref{efficiency}
which plots the beam intensity $\mathcal{I}$ thus deduced. These
results prove that the lithium pressure law can be safely
extrapolated,for $T \geq 673$ K.

On the same figure, we have plotted the beam intensity
$\mathcal{I}$  deduced from our measurements of the ion signal, as
if the detection probability was equal to $100\%$. These results
follow the same slope as the pressure law in this logarithmic
plot, further supporting the extrapolation of the pressure law.
During this experimental run, the rhenium ribbon temperature was
$T= 1590$ K and the residual gas pressure in the detector chamber
was $3\times 10^{-8}$ mbar (corresponding to $p_{O_2} \approx 6
\times 10^{-9}$ mbar if the residual gas is mostly air). From this
plot, we get the detection probability $D_+ = 32\pm 5 \%$. This
error bar does not include the remaining uncertainty on the
lithium vapor pressure.

\begin{figure}[htb]
\includegraphics{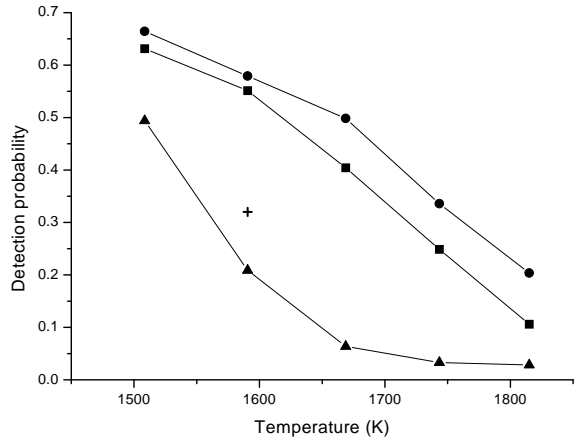}
\caption{\label{efficiency2} Measured detection probability as a
function of ribbon temperature for three different oxygen
pressures:  dots for $p_{O2} = 5.0 \times10^{-7}$ mbar, squares
for $p_{O2} = 2.1\times10^{-7}$ mbar, triangles for our best
vacuum ($3\times 10^{-8}$ mbar) corresponding to an estimated
oxygen pressure $p_{O2} =6\times10^{-9}$ mbar. From the
saturation of these curves, we estimate that the true ionization
probability is obtained by multiplying the detection probability
by $1/E_+= 1.41$ (see text). We have also represented by a cross
the value of $ D_+ = 32 \%$ at $T =1590$ K in our best vacuum
deduced from the data of figure \ref{efficiency}.}
\end{figure}

In a second series of experiments, we measured the detection
probability as a function of surface temperature and oxygen
pressure: the flux of lithium atoms impinging on the rhenium
ribbon was estimated to be $5.75\times 10^5$ atoms per second.
Before each measurement, the wire was flashed at $T= 2250$ K for
$2$ minutes, then the heating current was adjusted to reach the
working temperature and oxygen was admitted. After a stabilization
period, the ion signal was measured with the lithium beam on and
off and the background was substracted. We have varied the rhenium
ribbon temperature in the range $ 1590 - 1880 $ K. Two
measurements were made by introducing oxygen with a leak valve,
with oxygen pressures equal to $ 2.1\times10^{-7}$ and $5.0
\times10^{-7}$ mbar (uncorrected ion gauge readings) and one
measurement with our best vacuum ($3\times 10^{-8}$ mbar)
corresponding to $6\times10^{-9}$ mbar of oxygen, if the residual
gas is mostly air. We thus deduce the detection probability $D_+$,
which is plotted in figure \ref{efficiency2}. These new values of
the detection probability and the first measurement ($ D_+ = 32\pm
5 \%$ at $T =1590$ K in our best vacuum) are not in very good
agreement. This discrepancy, which illustrates the difficulty of
such absolute measurements, may be due to a variation of the
composition of the residual gas in the detection chamber (the
experiments corresponding to figure \ref{efficiency} have been
made one year before the experiments with a variable oxygen
pressure).

$D_+$ is the product of the ionization probability $P_+$ by an ion
counting efficiency $E_+$. This efficiency combines several
experimental factors (ion collection efficiency, channeltron
efficiency including electronic threshold effects) and all the
errors in the estimation of the atomic flux. We have no way of
measuring directly $E_+$, but the saturation of the detection
probability with a maximum observed value $D_+ =0.66 $ must
correspond to an ionization probability close to $100\%$,
following literature data
\cite{persky68,kawano86,kawano98,kawano00,kawano99}. From this
remark, we get a good estimate of the ion counting efficiency $E_+
= 0.71 $, this precise value being such that the corresponding
$P_+$ values depend smoothly on the rhenium surface oxidized
fraction $f$ defined in Appendix B. This data is analyzed in part
6.

The experiments of  Stienkemeier et al. \cite{stienkemeier00}
involve lithium atoms attached to helium droplets and the rhenium
ribbon is kept in a higher vacuum (residual pressure $5\times
10^{-9}$ mbar) than in our experiment. The efficiency was
estimated indirectly by comparison with other alkali atoms, with a
peak value of $10$ \% near $T= 1350 $ K. This value, substantially
lower than our own, corresponds to a lower degree of oxidation of
the rhenium surface, in agreement with the use of a higher vacuum.

\section{Optimization of the background signal}

The background signal has three main origins:
\begin{itemize}
\item
the components of the background gas with ionization potentials
below $9$ eV can be ionized on the rhenium surface
\item
a fresh rhenium wire contains a few ppm of alkali atoms. When the
wire is hot, these atoms diffuse inside the wire and reach the
surface where they are emitted as ions
\item
the background depends strongly on the oxidation of the rhenium
ribbon and this can be explained by the emission of rhenium oxide
ions.
\end{itemize}
\noindent We are going to discuss now these three sources of
background signals. All these contributions to the background
signal can be emitted from any point of the rhenium ribbon
surface, but the corresponding ions are detected only if they are
focused on the channeltron entrance funnel. A trick to reduce the
background signal is to limit ion collection to a small part of
the rhenium  ribbon. In our case, the ion optics collects the ions
emitted by the $5$ mm long central part of the ribbon and this
length cannot be much reduced, at least for our applications.

\subsection{Ionization of the background gas}

Surface ionization of many organic molecules is possible on
rhenium surfaces and this technique is well established, as
reviewed by Zandberg \cite{zandberg95}. The ionization probability
of any component of the residual gas depends rapidly on its
ionization potential $I$, through Saha-Langmuir law
(\ref{principle1}). For a species with a partial pressure close to
$10^{-8}$ mbar, about $5\times 10^{11}$ molecules impinge on a $1$
centimeter length of hot wire per second. Assuming a wire
temperature $T\simeq 1500K$ and an oxidized rhenium surface with
$\Phi \approx 6$ eV, the corresponding contribution to the
background remains below $10^2$ ions per second if $I-\Phi > 3$ eV
i.e. if $I \geq 9$ eV. Various types of organic molecules have
their first ionization potential near this value and can be
ionized: a mass spectrum produced by ionization of the residual
gas (produced by oil diffusion pumps) on an oxidized tungsten
ribbon is shown in ref. \cite{pauly68} and this spectrum is very
dense. It is therefore very important to operate in a high and
clean vacuum, because an oil-free vacuum contains mostly species
with high ionization potentials ($H_2$, $H_2O$, $CO$...).
Practically, the detector should be in a UHV chamber pumped by a
turbo molecular pump or an ion getter pump. A cold trap at liquid
nitrogen temperature may be useful to reduce the vapor pressure of
condensable species, but its use may not be necessary.

\subsection{Cleaning the alkali content of the rhenium ribbon}

Following Goodfellow \cite{goodfellow}, rhenium ribbons contain
$4$ ppm of potassium in mass corresponding to $3\times10^{14}$
potassium atoms per centimeter of ribbon length. At high
temperatures, these potassium atoms diffuse inside the wire, reach
its surface where they are emitted as ions. Using the diffusion
equation, we get the value of this ion current, assuming an
homogeneous initial potassium density and a wide ribbon ($a \ll
b$):

\begin{equation}
\label{background1}
 I_{ion} = \frac{8Nq}{\pi^2\tau_1}\sum_{p={\mbox  odd}}
\exp\left(-\frac{p^2t}{\tau_1} \right)
\end{equation}

\noindent where $N$ is the total number of potassium atoms in the
ribbon and $q$ is the modulus of the electron charge. This sum of
exponential is characteristic of a diffusion process, the
diffusion mode of order $p$ decaying with a time constant $\tau_p=
\tau_1/p^2$. For long times, the decay becomes purely exponential,
with the time constant $\tau_1$ related to the diffusion constant
$D$ by:

\begin{equation}
\label{background2}
 \tau_1 = \frac{a^2}{\pi^2 D}
\end{equation}

\begin{figure}[htb]
\includegraphics{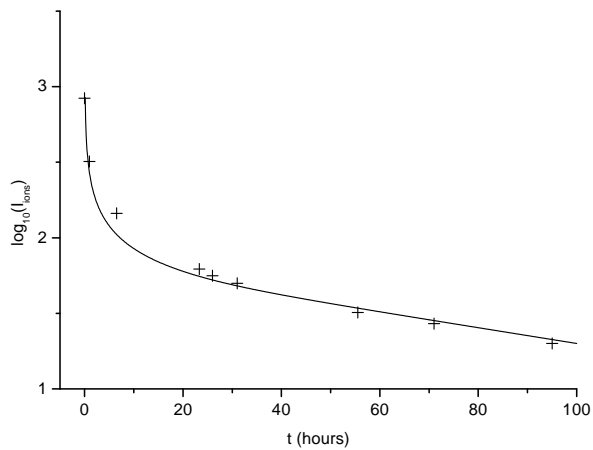}
\caption{\label{baking} Ion current in picoamperes (logarithmic
scale) as a function of time measured in hours (linear scale)
during the baking process at a temperature $T=2250$ K. The crosses
represent the measurements of the total ion current, while the
curve is the best fit to this data using equation
(\ref{background1}).}
\end{figure}

\noindent During the preparation of a ribbon made at a temperature
$T= 2250 $ K, the repeller plate (see figure 1) served to collect
about $38$ \% of the emitted ions. The corresponding total ion
current is plotted as a function of time in figure \ref{baking}.
These observations are well explained by equation
(\ref{background1}). From the longest time constant $\tau_1 \simeq
3.0\times 10^5$ s, we deduce the diffusion constant for potassium
inside rhenium $D = 0.3 \times 10^{-15}$ $m^2 s^{-1}$ at $T = 2250
$ K. As this time constant is long and as it scales like $a^2$, it
is very important to choose a ribbon with a small thickness $a$,
to minimize the duration of the cleaning process.

After $6.5$ hours of baking, we have briefly varied the wire
temperature $T$ and recorded the ion current as a function of
temperature. This current is proportional to the diffusion
constant, thus giving the temperature dependence of the diffusion
constant $D$, usually described by an Arrhenius law:

$$ D = D_0 \exp \left(-E_{diff}/k_BT \right)$$

\noindent From our measurements, we get the activation energy of
this diffusion process, $E_{diff}= 5.36\pm 0.23$ eV, and the
prefactor value, $D_0 = 0.3\times10^{-3}$ $m^2 s^{-1}$.

\subsection{Influence of rhenium oxidation on background signal}

While measuring the detection probability as a function of surface
temperature and oxygen pressure, we have recorded the background
signal. We discuss here only the data corresponding for $T= 1508$
K, because the temperature dependence is not very simple. With our
best vacuum (about $3 \times10^{-8}$ mbar), the background signal
was $1-3\times10^3$ counts per second and this signal increased
rapidly with oxygen pressure, reaching $1.6 \times 10^4$ and $
2.6\times 10^4$ counts per second  when the oxygen pressure was
$2.1\times10^{-7}$ or $5.0 \times10^{-7}$ mbar respectively.
Moreover, the noise of this background signal is substantially
larger than Poisson noise (up to $5$ times larger for $1$ second
counting periods). These two effects of a strong oxidation (rapid
increase of the background and excess noise) decrease considerably
the performance of the detector for a small atomic flux, even if
oxidation gives an important gain on detection probability.

\subsection{Operating conditions and ultimate performances}

We have already explained the need for a high and clean vacuum. If
one operates near $T=1500$ K, a small oxygen residual pressure of
the order of a few $10^{-9}$ mbar is sufficient to reach a
detection probability of the order of $10-30$ \% and to keep the
background at a very low level. If the initial baking time has
been sufficiently long, the potassium content of the rhenium
ribbon may be sufficiently reduced to give a small background
signal, as long as the working temperature is not too high. In
Bielefeld, where a very large set of experiments have been done
\cite{loesch93,hobel01}, and because the background pressure,
produced by an ion getter pump, is below $1\times10^{-8}$ mbar,
the typical background count rate is $100$ counts per second. In
Toulouse, where we have less experience and a less good vacuum
(near $3\times10^{-8}$ mbar produced by a turbo pump), the typical
background count rate is $1-3 \times 10^3$ counts per second.

Finally, in a high vacuum, surface contamination of the rhenium
ribbon by cracking of organic molecules is very slow, but it is
nevertheless necessary to clean the ribbon surface from time to
time by flashing it to a high temperature (typically $2250$ K) for
a few minutes, to get a reproducible operation.

\section{Time response of the detector: residence time of lithium
on a rhenium surface}

The time response of a detector is very important for many
applications. The dominant contribution for the present detector
is the ion residence time on the surface, usually in the
microsecond to millisecond range. The simplest measurement
technique uses a chopped atomic beam \cite{stienkemeier00}.
Another method, based on the autocorrelation function of the ion
current \cite{gladyszewski94}, has also  been be used and this
method extends the measurement range down to the microsecond
range, a value difficult to achieve by chopping atomic beams. All
the data sets present in the literature have been fitted by
equation (\ref{principle2}) namely $ \tau = \tau_0
\exp\left(E_{ads}/k_BT\right)$.

\begin{table}[htb]
\begin{center}
\begin{tabular}{|c|c|c|c|}
\hline
\ temperature (K) \ & \ $\tau_{exp}$ ($\mu$s) \ & \ $\tau$ ($\mu$s) \cite{stienkemeier00} \ & \ $\tau$ ($\mu$s) \cite{gladyszewski94} \ \\
\hline
1525 & 215 & 114  & 82 \\
\hline
1600 & 75 & 35 & 35 \\
\hline
\end{tabular}
\caption{Ion residence time as a function of the wire temperature.
Column 1 gives the temperature deduced from the heating
current (see appendix A), column 2 our measurement of the residence
time, columns 3 and 4 the calculated values of $\tau$ deduced from
ref. \cite{stienkemeier00} and \cite{gladyszewski94} respectively.
The data collected by these authors cover the temperature range
$1200-1500$ K (ref. \cite{stienkemeier00}) and $1600-2000$ K (ref.
\cite{gladyszewski94}) so that we have to extrapolate slightly
their fitted laws.}
\end{center}
\end{table}

Using a chopped atomic beam, we collected two measurements,
extending slightly the ranges covered by these works
\cite{stienkemeier00,gladyszewski94}. Table 1 presents a
comparison of our measurements to the values deduced from fitted
laws given in these two papers. The agreement between these
results is rather poor and this is easily understood for the
following reasons:

\begin{itemize}

\item  temperature measurement may be affected by systematic errors
(see appendix A) and any error on the temperature has a large
effect, because the residence time $\tau$ varies very rapidly with
temperature

\item in the simplest theoretical model (for example see ref.
\cite{scheer63b}), the adsorption energy increases with the work
function $\Phi$ and the work function depends on the degree of
oxidation of the rhenium surface. As different experiments test
rhenium surfaces with different degrees of oxidation, the
different values of the residence time for the same temperature
may be a real effect.

\end{itemize}

Table 2 collects the information concerning the residence time of
all the alkali on rhenium, so as to illustrate the trends followed
by $\tau_0$ and $E_{ads}$ through this chemical family. $\tau_0$
is expected to be close to the vibrational period of the ion near
the surface and this property, well verified in several cases, is
not verified by the data concerning lithium. The very long
extrapolation surely explains the scatter on $\tau_0$ with a
correlated modification of the adsorption energy (as an example,
the very different parameters of ref. \cite{stienkemeier00} and
\cite{gladyszewski94} for lithium lead to the same value the
residence time at $T = 1600$ K). The adsorption energy may also
vary with surface oxidation, as discussed above, but this effect
cannot explain the very large differences appearing in the case of
lithium .

\begin{table}[htb]
\begin{center}
\begin{tabular}{|c|c|c|c|}
\hline
\ reference & \ alkali & \ $\tau_0$ in $10^{-13}$ s & \ $E_{ads}$ in eV \ \\
\hline
\cite{scheer63a} & Cs & $1.9\pm 0.9$ & $2.01\pm 0.04$ \\
\hline
\cite{scheer63b} & Rb & $0.8\pm 0.3$ & $2.28\pm 0.03$ \\
\hline
\cite{scheer63b} & K & $1.0\pm 0.3$ & $2.33\pm 0.03$ \\
\hline
\cite{stienkemeier00} & K & $0.01\pm 0.06$ & $2.64\pm 0.06$ \\
\hline
\cite{scheer63b} & Na & $0.2\pm 0.1$ & $2.75\pm 0.03$ \\
\hline
\cite{gladyszewski94} & Na & $31.$ & $1.92$ \\
\hline
\cite{stienkemeier00} & Na & $0.05\pm 0.04$ & $2.95\pm 0.07$ \\
\hline
\cite{gladyszewski94} & Li & $12.$ & $2.37$ \\
\hline
\cite{stienkemeier00} & Li & $0.009\pm 0.006$ & $3.36\pm 0.05$ \\
\hline
\end{tabular}
\end{center}
\caption{Values of the preexponential factor $\tau_0$ and of the
adsorption energies $E_{ads}$ of the alkalis on rhenium. Column 1
gives the reference, column 2 the alkali, columns 3 and 4 the
values of $\tau_0$ and $E_{ads}$ respectively.}
\end{table}

\section{Modeling of the surface ionization process}

Surface ionization of a species $A$ is usually described using the
Saha-Langmuir equation (\ref{principle1}). This implicitly assumes
a thermal equilibrium between the desorbing particle and the
surface and is thus independent of the characteristics of the
electron transfer process between $A$ and the surface. As will be
shown below, it reproduces rather well the experimental
observations. However, one can wonder about the corresponding
microscopic view of the process and possible dynamical effects.
The charge transfer between an atomic projectile and a
free-electron metal surface is rather well understood
\cite{geerlings90} and quantitative descriptions are available
(see e.g. \cite{borisov96} for the alkali case). We can use this
microscopic description to predict the surface ionization
efficiency and compare it to the Saha-Langmuir prediction.

As an alkali atom approaches a metal surface, its electronic
levels couple with the continuum of metal states, resulting in a
finite lifetime of the atomic levels. The corresponding width
$\Gamma$ gives the charge transfer rate between the atom and the
surface. At $T = 0$ K, the electron is transferred from the atom
to the surface if the atomic level is above the surface Fermi
level and in the opposite direction if the atomic level is below
the Fermi level. For a finite temperature $T$, electron transfer
occurs in both directions according to the fractional population
of the metallic states at the atomic level energy position
\cite{geerlings90}. For a free-electron metal, the width of the
alkali level varies approximately exponentially with the distance
$z$ to the surface \cite{borisov96}. Qualitatively, when an alkali
leaves the surface, the very large width at small $z$ allows the
alkali charge state to reach thermal equilibrium; however, this is
not true at large $z$ and there exists a distance, called freezing
distance, $z_F$, beyond which the charge state of the desorbing
alkali decouples from the surface \cite{overbosch80}. Since the
alkali level energy is a function $E_a(z)$ (for the ionic state,
it roughly follows an image potential variation \cite{borisov96}),
the charge state equilibrium value changes with $z$ and the
asymptotic charge state of the desorbed particle is different from
its value on the surface and it a priori depends on the desorption
velocity. This 'freezing distance' discussion provides a
qualitative picture of the charge transfer between a projectile
and a metal surface. Quantitatively, if we assume that the
desorbing particle follows a classical trajectory $z(t)$, the
evolution of its charge state is governed by a rate equation
\cite{geerlings90}:

\begin{eqnarray}
\label{mod2} \frac{dn_+}{dt} &=& -2 \Gamma f(E_a(t),T) n_+(t) \nonumber \\
& &+\Gamma\left(1-n_+(t)\right)\left(1-f(E_a(t),T)\right)
\end{eqnarray}

\noindent where $n_+(t)$ is the ion charge fraction, $f(E_a(t),T)$
the Fermi function and $E_a(t)$ designates $E_a(z(t))$.

Surface ionization is modeled by solving equation (\ref{mod2}) for
a set of different classical trajectories, representing the
different possibilities for a desorbing alkali. The energies and
widths in (\ref{mod2}) are taken from the parameter-free
description of the alkali free-electron metal surface study of
ref. \cite{borisov96}. The initial state ($z = z_{ini}$) close to
the surface is taken ionic ($n_+=1$) and equation (\ref{mod2})
yields the survival probability of the ion at a large distance for
each trajectory and we then have to sum the contributions from the
different trajectories. In practice, we solve equation
(\ref{mod2}) up to a large $z$ distance for a set of total
energies, $E$, of the desorbing particle in the ionic channel. The
$z(t)$ trajectory introduced in equation (\ref{mod2}) is common to
the ionic and neutral desorption channels, which is not
appropriate for large $z$. To circumvent this problem in the
summation over the heavy particle energies, $E$, we assume that
the charge state stabilizes around the freezing distance $z_F$
where we compare the local kinetic energy of the desorbing
particle to the energy required for desorption in the ionic or
neutral channel in order to decide whether desorption is
energetically allowed or not. Ionic and neutral potential energy
curves being different, this leads to a lower desorption threshold
for neutral desorption. The different contributions are summed
with a thermal weighting factor to yield the surface ionization
probability $P_+$:

\begin{equation}
\label{mod3}\frac{ P_+}{1-P_+} = \frac{\int_0^\infty P_s^+(E)
\exp\left(-E/k_BT\right)dE}{\int_{E_{th}}^\infty \left(1-
P_s^+(E)\right) \exp \left(-E/k_BT \right)dE}
\end{equation}

\noindent where $E$ is measured with respect to the ionic
threshold and $E_{th}$ is equal to $E_{th} = E_a (\infty) -
E_a(z_F)$. The freezing distance, obtained from the level width
calculated in ref. \cite{borisov96}, is equal to $9.6$ $a_0$ for
lithium; it does not vary much in the energy range concerned in
surface ionization.

We thus get the surface ionization probability for a given surface
work function and temperature. Such an approach is indeed valid
for a free-electron metal and should also be meaningful in the
case of low adsorbate coverage of a metal. In the case of a low
adsorbate coverage on the surface, the perturbation of the charge
transfer is usually described in terms of local and non-local
effects (see a review on the adsorbate effects in
\cite{gauyacq98}). The local effect is due to the local potential
around the adsorbate which can strongly perturb the electron
transfer in a certain region surrounding the adsorbate. The
non-local effect comes from the surface work function change
induced by the adsorbate. An approach like the present one only
considers the non-local aspects. Local aspects are mostly visible
in scattering experiments which select specific trajectories, thus
probing specific areas on the surface. Surface ionization a priori
concerns the entire surface and thus should average over the local
effects. For very large adsorbate coverages, the electronic
structure of the surface is modified, possibly leading to an
insulator layer on the surface, on which the charge transfer
properties are different (see e.g. a review in \cite{borisov00}).

We are going to describe surface ionization of lithium on a
rhenium surface as a function of the temperature $T$ and of the
residual pressure, within the above approach and we will use the
rhenium work function extracted by Kawano et al. \cite{kawano99}
from their electron emission experiment (either directly the
published data or the modeling of these results presented here in
Appendix B). Figure \ref{theo1} presents the calculated surface
ionization probability of lithium for a residual air pressure of
$2.0\times 10^{-7}$ Torr, compared to the experimental results of
Kawano et al. \cite{kawano86}. The present microscopic results,
obtained with the surface work function modeling,  are seen to
reproduce the experimental trends rather well, the abrupt change
of the ionization probability from $100$\% down to a few \% as the
temperature is increased clearly appears to be connected with the
change of surface work function or equivalently to the degree of
oxidation of the rhenium surface. The two limits (high and low
$T$) in the present case correspond to either an almost clean
rhenium surface and an oxidized rhenium surface. The temperature
$T_c$ at which the abrupt change occurs depends on the residual
gas pressure which directly influences the oxygen adsorption
change with temperature. $T_c$ decreases when the residual
pressure decreases. Typically, it changes by around $300$ K  for a
residual air pressure change between $10^{-9}$ and $2\times
10^{-6}$ Torr. The experimental and theoretical variations of the
ionization probability are shifted one with respect to the other
by around $40$ K. This can be due to the approximations involved
in the present modeling and/or to the accuracy of the temperature
scale (see below and Appendix A). Finally, the asymptotic value of
the ionisation probabibility at high temperature is underestimated
by our calculation and this proves that we use a too low value of
the work function $\Phi_0$ of clean unoxidized rhenium. Clearly,
this value, $\Phi_0 = 4.94$ eV, taken from our fit of Kawano et
al. \cite{kawano99} data ( see Appendix B), is slightly lower
than, for instance, the value obtained by Persky \cite{persky68},
$\Phi_0 = 5.0 \pm 0.1 $ eV.

\begin{figure}[htb]
\includegraphics{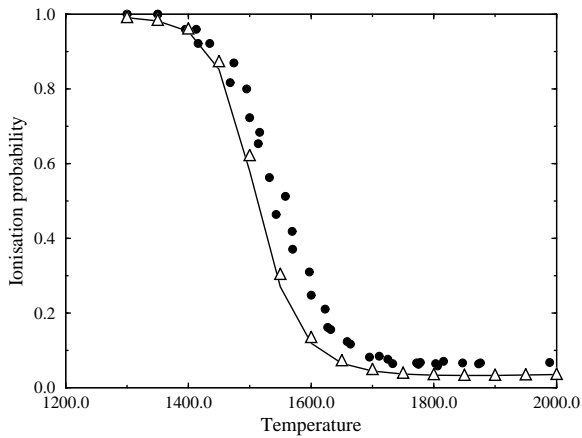}
\caption{\label{theo1} Ionization probability of lithium on
rhenium as a function of surface temperature with an air pressure
of $2\times 10^{-7}$ Torr: the full curve is the Saha-Langmuir
law, using our modeling of the work function (Appendix B); the
empty triangles represent the theoretical calculation described in
part 6, with the same value of the work function; the dots
represent the data of Kawano et al. \cite{kawano86} }
\end{figure}

For surface ionization of sodium on rhenium surface, a similar
agreement (not shown here) is found between the present modeling
and the experimental results of Kawano et al. \cite{kawano98}.
Figure \ref{theo1} also presents the prediction of the
Saha-Langmuir equation for the same conditions. The predictions of
the present microscopic study are extremely close to those of the
Saha-Langmuir equation, typically within a couple of \%. In fact,
this can be understood if we replace the value of $P_s^+$ from the
numerical solution of equation (\ref{mod2}) by its freezing
distance approximation \cite{overbosch80}, i.e. if the final
charge state $P_s^+$ is taken equal to its equilibrium value at
the freezing distance:

\begin{equation}
\label{mod4}
 P_s^+ = \frac{1}{1 + 2 \exp(\frac{E_a(z_F)-\Phi}{k_B T})}
\end{equation}

\noindent Bringing equation (\ref{mod4}) into equation
(\ref{mod3}) then leads to the Saha-Langmuir equation
(\ref{principle1}), the value of $E_a(z_F)$ disappearing from the
result. One can notice that the freezing distance approximation
(\ref{mod4}) consists in applying the Saha-Langmuir law expression
for the electronic levels alone at the point where the electronic
levels decouple; the sum over the heavy particle energies
(equation (\ref{mod3})), which takes into account the energy
changes between $z_F$ and infinity, transforms it into the usual
Saha-Langmuir expression for the total energy of the system
evaluated at infinity. The good agreement between the present
microscopic model and Saha-Langmuir expression thus proves that
dynamical effects are absent and that the specificity of the metal
surface-projectile charge transfer process disappears. This
comforts the validity of a thermal equilibrium approach for the
surface ionization process.

\begin{figure}[htb]
\includegraphics{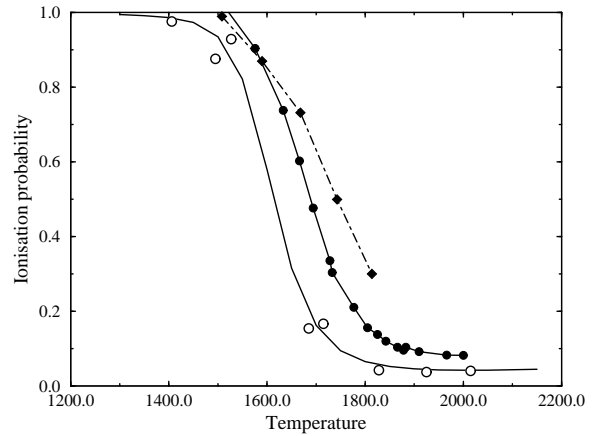}
\caption{\label{theo2} Ionization probability of lithium on
rhenium as a function of surface temperature with an air pressure
of $2\times 10^{-6}$ Torr: the full curve is the our theoretical
modeling of the ionization process described in part 6, using the
equations of Appendix B to describe the work function; the empty
circles represent this same theoretical modeling calculation, but
using directly the measurements of work function \cite{kawano99};
the dots represent the data of Kawano et al. \cite{kawano86} and
the diamonds our results (shown in figure \ref{efficiency2})
multiplied by a factor $1/E_+= 1.41$ to get the true ionization
probability. }
\end{figure}

Figure \ref{theo2} presents a comparison of experimental and
theoretical results for the surface ionization probability for a
residual air pressure of $2.0\times 10^{-6}$ Torr.  Two
theoretical results are presented which were obtained either by
using directly the surface work function extracted from electron
emission experiments \cite{kawano99} or by using its modeling
(Appendix B). The two results are close one to the other,
confirming the efficiency of the modeling of the work function
change. The two experimental sets are the results from reference
\cite{kawano86} and the present results. The latter have been
multiplied by $1.41= 1/E_+$ to transform our detection probability
result $D_+$ into an ionization probability $P_+$. Although the
general behaviour of $P_+$ as a function of T is the same in the
three sets, they appear shifted one with respect to the other. The
abrupt change in $P_+$ as a function of T from ref.
\cite{kawano86} is more rapid than in the present study. In both
cases, there is an upward temperature shift when going from the
model to the experiments (about $70$ K for the data of ref.
\cite{kawano86} and $110$ K for our results). These differences
are tentatively attributed to the accuracy of the temperature
scales, both in the experimental results on surface ionization
(estimated to $5$ \% in the present study) and in the modeling via
the use of the electron emission experimental results
\cite{kawano99}.

\section{Acknowledgements}
We want to thank  J. P. Toennies, J. P. Ziesel, A.
Bordenave-Montesquieu for their help and advice and P. Echegut for
information on rhenium emittance and advice on temperature
measurements. We thank H. Kawano, L. Gladyszewski, F. Stienkemeier
and M. Wewer for helpful correspondence. R\'egion Midi
Pyr\'en\'ees is gratefully acknowledged for financial support.

\section*{Appendix A: Hot wire temperature}

For a long ribbon of thickness $a$ and width $b$, the equilibrium
temperature $T$ results from the equilibrium between the input
power due to Joule effect and radiative losses:
\begin{equation}
\label{A1}
 \frac{\rho (T) I^2}{a b} = \epsilon(T) \sigma T^4 [2(a
+ b)]
\end{equation}

\noindent where $\rho (T)$ is the electrical resistivity,
$\epsilon(T)$ the total emittance, both temperature dependent and
$\sigma$ the Stefan-Boltzmann constant. It is a very good
approximation to forget thermal conduction along the wire, as long
as one does not want to describe the temperature distribution near
the ribbon ends. Rhenium resistivity in $\Omega \cdot m$
\cite{rembar} is well fitted by:
\begin{equation}
\label{A2}
 \rho(T) = 26.0 \times 10^{-8} \times(1 +
1.27\times 10^{-3}\times T)
\end{equation}
\noindent for $1200 <T<2000 $ K. For the total emittance of
rhenium, we have also fitted the three data sets collected in ref.
\cite{touloukian70} and covering the range $300 <T<3000$ K by a
linear function of $T$:

\begin{equation}
\label{A3}
 \epsilon(T) = 0.0852 \times (1 + 1.15\times 10^{-3}\times T)
\end{equation}

\noindent all the data points being within $\pm 20\%$ of this fit:
this uncertainty produces the dominant temperature error bar equal
to $5\%$. It appears from equations (\ref{A2},\ref{A3}) that the
ratio $\rho(T)/\epsilon(T)$ is a  very slow function of the
temperature $T$. This explains why the simple law $T \propto
\sqrt{I}$ is an excellent approximation as noted by ref.
\cite{stienkemeier00}. In the range $1200-2000$ K, we get:

\begin{equation}
\label{A4} T = 1124\times \sqrt{I}
\end{equation}

\noindent where $I$ is measured in ampere. The numerical factor
$1124 $ corresponds to the dimensions ($a= 30$ $\mu m$ and $b=760$
$\mu m$) of the ribbon used in Toulouse and, for other ribbon
geometries, this factor is easily scaled, using equation
(\ref{A1}). All temperatures appearing in this paper were deduced
from this equation (even up to $2250$ K), without recalling our
estimated $5$\% error bar.

We have also used an optical pyrometer to measure the temperature:
the sensitivity is close $\pm 5$ Kelvin, but we have no way to
test its calibration. Moreover, such pyrometers are calibrated to
measure the temperature of a blackbody radiation and the readings
$T_r$ must be corrected to take into account rhenium spectral
emittance \cite{touloukian70} near $655$ nm $\epsilon_{655} \simeq
0.40$ and window transmission $\theta$. The corrected temperature
$T_c$ is given by:

\begin{equation}
\label{A5} \frac{1}{T_c} = \frac{1}{T_r} + \frac{k_B}{h
\nu_{655}} \ln(\epsilon_{655}\theta)
\end{equation}

\noindent This correction is substantial (about $200$ K near
$2000$ K). With our pyrometer, the readings $T_r$ are lower than
the values of $T$ deduced from equation (\ref{A4}) by about $5\%$
while the corrected values $T_c$ are higher by roughly the same
amount.

\section*{Appendix B: Rhenium work function $\Phi$ as a function
of oxidation and temperature}

The strong influence of oxidation of rhenium surface on the
ionization probability of lithium was first observed by Persky
\cite{persky68} in 1968. This study was continued by Kawano and
coworkers \cite{kawano86,kawano98,kawano00}, who used the
Saha-Langmuir law to deduce the work function from the observed
ionization probability. In 1999, Kawano et al. \cite{kawano99}
also measured the work function of oxidized rhenium from the
emitted electron current, using the Richardson law ( $J = A T^2
\exp\left( -\Phi/k_BT\right)$) to relate the current density $J$
to the work function $\Phi$. The two values of the work function
differ noticeably, the work function for positive ion emission
deduced from the Saha-Langmuir law being substantially larger than
the work function for electron emission deduced from the
Richardson law.

We first discuss the work function for electron emission because
Kawano and coworkers \cite{kawano99} have collected a large data
set as a function of oxygen pressure and rhenium temperature. We
have developed a simple model which fits these results very
satisfactorily. The oxygen surface coverage is assumed to be
described by the following chemical equilibrium:
\begin{equation}
\label{oxy1} 2 \mbox {  rhenium sites   } + O_2 \rightleftharpoons
2 \mbox {  oxidized rhenium sites}
\end{equation}
Let $\sigma_{Re}$ and $\sigma_{ReO}$ be the density of unoxidized
and oxidized sites respectively, and $p(O_2)$ the partial pressure
of molecular oxygen. The equation resulting from this chemical
equilibrium is:
\begin{equation}
\label{oxy2}
 \left(\frac{\sigma_{ReO}}{\sigma_{Re}}\right)^2 = \frac{p(O_2)}{p_c(T)}
\end{equation}

\noindent where $p_c(T)$ is the equilibrium constant of the
reaction, written so as to have the dimension of a pressure. From
this equation, we can deduce the fraction $f$ of the oxidized
sites:
\begin{equation}
\label{oxy3}
 f= \frac{\sigma_{ReO}}{\sigma_{Re}+ \sigma_{ReO}} =
 \frac{\sqrt {p(O_2)/p_c(T)}}{1+ \sqrt {p(O_2)/p_c(T)}}
\end{equation}

\noindent We would get a different equation if we consider that
each rhenium site accepts two oxygen atoms. We have also tried to
fit the data with this modified equation. As the resulting fit is
considerably less good, we consider that this second hypothesis is
not correct. The second assumption of the model is that the work
function $\Phi$ increases linearly with the coverage $f$ from the
pure metal value $\Phi_0$ to the completely oxidized value noted
$\Phi_0 +\Delta$:
\begin{equation}
\label{oxy4}
 \Phi = \Phi_0 + f \Delta
\end{equation}

\begin{figure}[htb]
\includegraphics{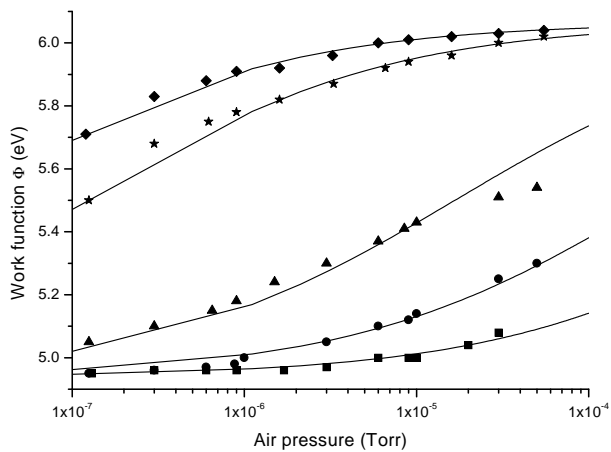}
\caption{\label{workfunction} Work function of rhenium as a
function of air pressure in Torr (logarithmic scale) for various
temperatures: diamonds for $T= 1406$ K, stars for $T= 1527$ K,
triangles for $T= 1685$ K, dots for $T= 1828$ K and squares for
$T=2015$ K. The experimental results are taken from figure 4 of
the paper by Kawano et al. \cite{kawano99}, while the curves
represent our best fit to these data, using equations (\ref{oxy3})
and (\ref{oxy4}).}
\end{figure}

\noindent This assumption is surely oversimplified but we do not
see how to refine easily this model as it already represents quite
well the results presented of H. Kawano et al. \cite{kawano99} in
the temperature range $1400-2000$ K. Our fit to these results is
presented in figure \ref{workfunction}. Below $1400$ K, the
observed variations of the work function deviate strongly from the
trend observed at higher temperatures and we have not tried to fit
this data. From the fits, we extract the following values: $\Phi_0
= 4.94$ eV, $\Delta = 1.125$ eV and for each temperature a value
of the equilibrium constant $p_c(T)$. The data of figure 5 of
reference \cite{kawano99} are well fitted too, but with a slightly
lower $\Delta$ value, $\Delta = 1.00$ eV. From these fits, we have
deduced 8 values of $p_c(T)$. A good test of the consistency of
our model is that, as expected from thermodynamics, these values
are well represented by the following formula:

\begin{equation}
\label{oxy5} log_{10}(p_c(T)) = A+\frac{B}{T}
\end{equation}

\noindent If the pressures are expressed in Torr, we get $A=
9.186$ and $B= -25 072$ K. The quantity $B$ corresponds to an
energy $B= - 2.16$ eV, which represents the energy difference
between the two sides of the chemical equilibrium described by
equation (\ref{oxy1}). This modeling provides a very efficient way
of interpolating between the measured work function values and we
used it in our treatment of the surface ionization to define the
surface work function as a function of the operating conditions.

The work function of rhenium can also be deduced from the
measurements of the ionization probability of sodium or lithium
\cite{kawano86,kawano98,kawano00}. The analysis is somewhat more
complex because the experiments were done with intense alkali
halide beams and the dissociation equilibrium of the molecule on
the surface must be taken into account. The Saha-Langmuir law is
used to extract the work function $\Phi_+$ (thus noted as it
differs from the electron emission value) from the data as a
function of the surface temperature and oxygen pressure. However,
two aspects of the analysis of Kawano and coworkers  deserve
further discussion:

\begin{itemize}
\item the ionization potential $I$ of the alkali atom is taken as
a function of the wire temperature (this function appears to be
$I(T) = I_0 + 2.5 k_BT$). This is a substantial effect which is
not commonly considered (see for instance ref. \cite{kaack95}, in
which very accurate measurements of the platinum and iridium work
function are made as a function of the temperature). Adding this
unusual T dependence to the atomic ionization energy significantly
contributes to the observed difference between the surface work
function extracted from electron emission and the effective work
function extracted from surface ionization data.

\item it is surely very difficult to estimate the ionization
probability to better than a few percent, in particular because
the flux incident on the wire is calculated from vapor pressure
data which are not very accurate (as discussed in the present
paper in the lithium case). It is therefore very difficult to
understand how the work function $\Phi_+$ can reach a value as
large as $6.7$ eV (see figure 3b of ref. \cite{kawano98} where the
experiment was done with NaCl): as soon as $(\Phi_+ -I) \geq 5.3
k_B T $, the Saha-Langmuir law predicts $P_+ \geq 0.99$, almost
impossible to distinguish experimentally from $P_+ =1$. For
temperatures close to $1500$ K, the maximum value of $\Phi_+$,
which can be reliably deduced from Saha-Langmuir law, is not
larger than $6.1$ eV when working with lithium (and only $5.8$ eV
when working with sodium for which $I= 5.139$ eV).

\item finally, one can stress that most of the experimental results of
ref. \cite{kawano86} and \cite{kawano98} on the surface ionization
probability of lithium and sodium can be reproduced by
Saha-Langmuir law using the surface work function extracted from
electron emission \cite{kawano99} and the ionization potential of
the free atom. There does not seem to be any need to introduce an
effective work function for surface ionization, which could be, at
best, only a parameterization of the experimental results.
\end{itemize}

Therefore, we think that the experimental results on $P_+$,
obtained by Kawano and co-workers are very interesting for the
characterisation of a surface ionization detector, but the values
of $\Phi_+$ extracted from these should be considered with
caution.



\end{document}